\documentclass[aps,preprint,superscriptaddress,showpacs,floatfix]{revtex4}

\usepackage{colortbl}
\usepackage{graphicx}
\usepackage{dcolumn}
\usepackage{bm}
\usepackage{CJK}
\usepackage{amsmath,amssymb,latexsym}

\begin{document}

\title{Magnetic rotations in $^{198}\rm Pb$ and $^{199}\rm Pb$ within covariant density functional theory}

\author{L. F. Yu}
\affiliation{State Key Laboratory of Nuclear Physics and Technology, School of Physics, Peking University, Beijing 100871, China}%
\author{P. W. Zhao}
\affiliation{State Key Laboratory of Nuclear Physics and Technology, School of Physics, Peking University, Beijing 100871, China}%
\author{S. Q. Zhang}\email{sqzhang@pku.edu.cn}
\affiliation{State Key Laboratory of Nuclear Physics and Technology, School of Physics, Peking University, Beijing 100871, China}%
\author{P. Ring}
\affiliation{State Key Laboratory of Nuclear Physics and Technology, School of Physics, Peking University, Beijing 100871, China}%
\affiliation{Physik Department, Technische Universit\"at M\"unchen, D-85748 Garching, Germany}%
\author{J. Meng}\email{mengj@pku.edu.cn}
\affiliation{State Key Laboratory of Nuclear Physics and Technology, School of Physics, Peking University, Beijing 100871, China}%
\affiliation{School of Physics and Nuclear Energy Engineering, Beihang University, Beijing 100191, China}%
\affiliation{Department of Physics, University of Stellenbosch, Stellenbosch, South Africa}%

\begin{abstract}
Well-known examples of shears bands in the nuclei $^{198}\rm Pb$ and $^{199}\rm Pb$ are investigated within tilted axis cranking relativistic mean-field theory.
Energy spectra, the relation between spin and rotational frequency, deformation parameters and reduced $M1$ and $E2$ transition probabilities are calculated.
The results are in good agreement with available data and with calculations based on the phenomenological pairing plus-quadrupole-quadrupole tilted-axis
cranking model. It is shown that covariant density functional theory provides a successful microscopic and fully self-consistent description of
magnetic rotation in the Pb region showing the characteristic properties as the shears mechanism and relatively large $B(M1)$ transitions decreasing
with increasing spin.
\end{abstract}
\maketitle


\section{Introduction}

Magnetic rotation has become one of the most important high-spin phenomena in atomic nuclei during the past two decades. The special interest in magnetic rotational bands arises mainly from their unusual properties that differ from conventional collective rotational bands in well-deformed nuclei. Magnetic rotational bands observed in nearly spherical or only weakly deformed nuclei have very weak and sometimes unobservable electric $E2$ transitions but rather strong magnetic $M1$ transitions. The explanation of such bands was given in Ref.~\cite{Frauendorf1993Nucl.Phys.} in terms of the shears mechanism. In a classical picture, in these bands the magnetic dipole vector arising from proton particles (holes) and neutron holes (particles) in high-$j$ orbitals rotates around the total angular-momentum vector. The name ``magnetic rotation'' was introduced~\cite{Frauendorf1994,Frauendorf1996} to account for such a kind of rotation and was discussed in detail on the general basis of spontaneous symmetry breaking in Refs.~\cite{Frauendorf1997Z.Phys.A,RMP73(2001)463}. Meanwhile, with increasing spin, the particles and holes in the high-$j$ orbitals align along the total angular momentum and this alignment reduces the perpendicular component of the magnetic dipole moment. As a result, a typical property of these shears bands is the decreasing of the $B(M1)$ values with increasing spin.

Magnetic rotational bands have been observed in several mass regions with the mass numbers $A\sim60$, $A\sim80$, $A\sim110$, $A\sim140$, and $A\sim190$~\cite{Amita2000AtomicDataandNuclearDataTables,Torres2008Phys.Rev.C}, but the richest information exists for neutron-deficient Pb isotopes where the coupling between the proton $1h_{9/2}$ and $1i_{13/2}$ particles and the neutron $1i_{13/2}^{-n}$ holes provides the necessary particle-hole configurations. In particular, long cascades of $M1$ transitions were observed first in neutron deficient Pb nuclei in the early 1990s~\cite{Clark1992Phys.Lett., Baldsiefen1992Phys.Lett., Kuhnert1992Phys.Rev.C}. Later, in 1997, the lifetime measurements based on the Doppler-shift attenuation method (DSAM) for four $M1$ bands in the nuclei $^{198}\rm Pb$ and $^{199}\rm Pb$  provided a clear evidence for magnetic rotation~\cite{Clark1997Phys.Rev.Lett.}. Subsequently, another  experiment using the recoil distance method (RDM) in $^{198}\rm Pb$ together with the results of the DSAM experiment provided further support to the shears mechanism~\cite{PRC58(1998)1876}. Naturally, magnetic rotation in the Pb region deserves great theoretical efforts.

On the theoretical side, tilted axis cranking (TAC) mean-field models~\cite{Frauendorf1993Nucl.Phys.,Frauendorf2000Nucl.Phys.,RMP73(2001)463} have been the main tools for studying the magnetic dipole bands. In addition, the shell model~\cite{Frauendorf1996Nucl.Phys.} and the many-particles-plus-rotor model~\cite{PhysRevC.74.044310} have also been applied to investigate magnetic rotation in the Pb region. So far, the TAC models have turned out to be powerful tools to describe magnetic rotational bands, because these models are based on a classical picture of rotation and here it is relatively easy to construct vector diagrams showing the angular-momentum composition.
This method is of great help to visualize the structure of magnetic rotation in these bands. In Ref.~\cite{Frauendorf1996ZP} the quality of the TAC approximation has been discussed and tested in comparison with the particle rotor model. Similar investigations along this line have been extended for three-dimensional cranking and the nuclear chirality is proposed~\cite{Frauendorf1997NP}. Because of the high numerical complexity of the TAC model, even today most of the applications are based on simple phenomenological  Hamiltonians, such as the pairing plus-quadrupole-quadrupole tilted-axis
cranking (PQTAC) model~\cite{Chmel2007Phys.Rev.C}. Only recently nonrelativistic~\cite{Olbratowski2002Acta.Phys.Pol.B} and relativistic~\cite{Madokoro2000Phys.Rev.C,Peng2008Phys.Rev.C, Zhao2011181} codes have been developed for microscopic investigations of magnetic rotation in the framework of density functional theory. By obvious reasons they have first been applied in lighter nuclei with smaller model spaces. So far there exists no microscopic investigation in the Pb region.

During the past several decades, covariant density functional theory (CDFT) has attracted wide attention due to its success in describing many phenomena in stable as well as exotic nuclei~\cite{Ring1996Prog.Part.Nucl.Phys., Vretenar2005101, Meng2006Prog.Part.Nucl.Phys.}. It exploits Lorentz invariance, one of the underlying symmetries of QCD. This symmetry allows to describe in a consistent way the spin-orbit coupling having an essential influence on the underlying shell structure in finite nuclei. It also puts stringent restrictions on the number of parameters in the corresponding functionals without reducing the quality of the agreement with experimental data. CDFT with meson exchange presents a microscopic treatment of the nuclear many-body problem in terms of nucleons and mesons carrying the effective interaction between nucleons. Moreover, since the theory is relativistically invariant and the field and nucleon equations of motion are solved self-consistently, such models preserve causality and provide a self-consistent description of the spin-orbit term of the nuclear effective force and of the bulk and surface parts of the interaction. For the description of magnetic rotations it is of particular importance that these functionals include {\it nuclear magnetism}~\cite{KR.89}, i.e., a consistent description of currents and time-odd fields. This plays a role for odd-mass nuclei~\cite{AA.10}, excitations with unsaturated spins, magnetic moments~\cite{HR.88,PRC74(2006)024307}, and nuclear rotations~\cite{ARK.00}. As a consequence of Lorentz invariance no new parameters are required for the time-odd parts of the mean fields. In nonrelativistic functionals the corresponding time-odd parts are usually difficult to adjust to experimental data and even if there are additional constraints derived from Galilean invariance and gauge symmetry~\cite{DD.95} these constraints are usually not taken into account in the successful functionals commonly used in the literature.

The earlier versions of CDFT were based on the Walecka model~\cite{DT.56,Due.56,Wal.74,SW.86} with phenomenological nonlinear meson-interactions proposed by Boguta and Bodmer~\cite{BB.77} introducing in this way a phenomenological density dependence~\cite{NL3,TP.05,PK1}. The nonlinear models have been replaced by an explicit density dependence of the meson-nucleon vertices~\cite{BT.92}. For successful functionals of this type this density dependence has been determined in a phenomenological way~\cite{TW.99,DD-ME1,DD-ME2} and such models have shown considerable improvements with respect to previous relativistic mean-field (RMF) models in the description of asymmetric nuclear matter, neutron matter, and nuclei far from the valley of stability. On the other hand, on a more microscopic way, one has tried to derive the density dependence from Brueckner calculations in nuclear matter at various densities~\cite{BT.92,HKL.01,SOA.05,HSR.07}. An example is density dependent relativistic hadron field theory~\cite{HKL.01} where the specific density dependence of the meson-nucleon vertices is mapped from Dirac-Brueckner calculations where the in-medium interaction is obtained from nucleon-nucleon potentials consistent with scattering experiments. Therefore, if this ansatz is adopted, the effective theory is derived fully from {\it ab initio} calculations. Of course, the accuracy of the results obtained in this way is by no means satisfactory for modern nuclear structure calculations and a fine tuning of additional free parameters is still needed. This fact allows us to constrain the different possibilities and keeps the compatibility, at least theoretically, with more fundamental calculations of infinite nuclear matter.

As Inglis~\cite{Ing.56} has shown already in the 1950s, rotational excitations in nuclei can be described in the framework of the cranking model, i.e., in a uniformly rotating mean field, where the rotational frequency $\Omega$ is determined in a semiclassical way by the condition $\langle J_{x} \rangle =\sqrt{I(I+1)}$.  This model has been extended to a very successful self-consistent description of rotational nuclei all over the periodic table; first, with simple phenomenological quadrupole-pairing forces~\cite{RBM.70,RM.74} and later with interactions derived from nonrelativistic~\cite{FMR.79,BFH.87,ERo.93,GDB.94} and relativistic~\cite{KR.89,AKR.96} density functionals. It also has been shown that the cranking wave functions correspond to intrinsic wave functions obtained by a variation after projection onto good angular momentum~\cite{BMR.70}.  Most of  these calculations have been carried out for collective rotations with a rotational axis perpendicular to the symmetry axis. This method is called principal axis cranking (PAC) and also {\it one-dimensional cranking}, because there is only one component of the angular velocity of importance. The model has been extended to rotations around an arbitrary axis tilted against the principal axis of the density distribution~\cite{Bengtsson1987, Frauendorf1997NP,Madokoro2000Phys.Rev.C,Olbratowski2004Phys.Rev.Lett.}, i.e., {\it three-dimensional cranking}. In the framework of CDFT, a code for the solution of these equations has been developed in Ref.~\cite{Madokoro2000Phys.Rev.C}. However, because of the numerical complexity, it has only been applied for the magnetic rotation in $^{84}\rm Rb$ a relatively light nucleus. Later on, focusing on magnetic rotation, a {\it two-dimensional cranking} version of RMF model based on the nonlinear meson-exchange interaction has been established in Ref.~\cite{Peng2008Phys.Rev.C}. It has been applied to the medium heavy nucleus $^{142}\rm Gd$. Recently, a two-dimensional tilted axis cranking relativistic mean-field (TAC-RMF) version based on the point-coupling interaction~\cite{PRC46(1992)1757, Zhao2010Phys.Rev.C} has been developed.  The zero range of the corresponding effective interaction simplifies the code considerably allowing systematic investigations~\cite{Zhao2011181,Zhao2011PRL}.

In the present work, this self-consistent TAC-RMF model based on a point-coupling interaction is applied to the $A\sim190$ region for the first time. The magnetic rotational bands in $^{198,199}$Pb are investigated. The energy spectra, the relation between spin and rotational frequency, deformation parameters and reduced $M1$ and $E2$ transition probabilities are calculated and compared with the available data~\cite{NPA683(2001)108,EPJA5(1999)257,Clark1997Phys.Rev.Lett.,PRC58(1998)1876} and those obtained from the PQTAC model~\cite{Chmel2007Phys.Rev.C}.

The paper is organized as follows: after establishing the basics of relativistic point-coupling models in Sec.~\ref{section1}, we discuss in Sec.~\ref{section2} numerical details of the method. In Sec.~\ref{section3} we compare the results of these calculations with experimental data. Section~\ref{conclusion} contains conclusions and an outlook for future work.

\section{THEORETICAL FRAMEWORK}%
\label{section1}
The starting point of CDFT is an effective Lagrangian density from which the equations of motion can be obtained. For a nucleus with$D_ {2}$ symmetry, the TAC-RMF model assumes that the nucleus rotates around an axis in the $xz$ plane of the body-fixed system. Assuming a rotation with constant velocity around a fixed axis in space the effective Lagrangian is transformed into the rotating frame~\cite{KR.89}. In this system one is confronted with a quasistationary problem and one obtains the following equations of motion:
\begin{equation}
[\bm\alpha\cdot(-i\nabla-{\bm V})+\beta(m+S)+V-\bm{\mathit\Omega}\cdot\bm {\hat J}]|\psi_i\rangle=\epsilon_i|\psi_i\rangle,
\label{dirac}
\end{equation}
where $\bm{\mathit\Omega}= (\mathit\Omega_x, 0, \mathit\Omega_z) = (\mathit\Omega\cos\theta_{\mathit\Omega}, 0, \mathit\Omega\sin\theta_{\mathit\Omega})$ is the rotational frequency with the tilted angle $\theta_{\mathit\Omega}\mathrel{\mathop :}=\sphericalangle(\bm{\mathit\Omega}, \bm e_x)$ between the cranking axis and the $x$ axis , $\bm{\hat J}=\bm{\hat L}+\frac{1}{2}\bm{\mathit{\hat\Sigma}}$, is the total angular momentum of the nucleus, the sum of all the single-particle angular momenta. Using the point-coupling model discussed in  Ref.~\cite{Zhao2010Phys.Rev.C} the relativistic fields have the form
\begin{eqnarray}
S(\bm{r})&=&\alpha_S\rho_S+\beta_S\rho_S^2+\gamma_S\rho_S^3+\delta_S\triangle\rho_S,\\
V^\mu(\bm{r})&=&\alpha_V j_V^\mu+\gamma_V(j_V^\mu)^3+\delta_V\triangle j_V^\mu\\
&+&\tau_3\alpha_{TV} j_{TV}^\mu+\tau_3\delta_{TV}\triangle j_{TV}^\mu+e\frac{1-\tau_3}{2}A^\mu,
\nonumber
\end{eqnarray}
where $\alpha_S, \beta_S, \gamma_S, \delta_S, \alpha_V, \gamma_V, \delta_V, \alpha_{TV}$, and $\delta_{TV}$ are the coupling constants. $\alpha$ refers to the four-fermion term, $\beta$ and $\gamma$ to the third and the four-order terms, and  $\delta$ to the derivative couplings. The subscripts $S$, $V$, and $T$ stand for scalar, vector, and isovector couplings, respectively. The iterative solution of the equations of motion yields single-particle energies, energy, quadrupole moments, magnetic moment, and so on. More details can be found in Refs.~\cite{Peng2008Phys.Rev.C, Zhao2011181}.

The size of the rotational frequency $\Omega$ is connected to the total angular-momentum quantum number $I$ by the semiclassical relation
$J^2=\langle\hat J_x\rangle^2+\langle\hat J_z\rangle^2=I(I+1)$, and the orientation of rotational frequency vector $\bm\Omega$ is determined by minimizing the total Routhian
$\langle \hat H-\bm{\mathit\Omega}\cdot\bm{\hat J}\rangle$ for fixed orientation of the angular momentum $\bm J$ and for a fixed absolute value of the angular velocity
$|\Omega |$. It turns out that this requirement leads to a parallel alignment of the two vectors $\bm{\Omega}$ and $\bm J$~\cite{Peng2008Phys.Rev.C}.

The nuclear magnetic moment is derived as an expectation value of the relativistic form of the effective current operator
   \begin{equation}%
      \label{magneticmoment}
         \bm{\mu}=\sum_{i=1}^A\langle\psi_i|\frac{mc^2}{\hbar c}q\,\bm{r}\times\bm{\alpha} +\kappa\,\beta\,\bm{\Sigma}|\psi_i\rangle,
   \end{equation}
where $q$ is the charge ($q_p=1$ for protons and $q_n=0$ for neutrons), $m$ the nucleon mass, and $\kappa$ the free anomalous gyromagnetic ratio of the nucleon ($\kappa_p=1.793$ and $\kappa_n=-1.913$). In a semiclassical approximation, the $B(M1)$ values are~\cite{Frauendorf1993Nucl.Phys.,Frauendorf1997NP}
\begin{equation}
B(M1)=\frac{3}{8\pi}\mu_{\bot}^2=\frac{3}{8\pi}(\mu_x\sin\theta_{\mathit\Omega}-\mu_z\cos\theta_{\mathit\Omega})^2.
\end{equation}

\section{NUMERICAL DETAILS}
\label{section2}

\begin{figure*}
\centerline{
\includegraphics[scale=0.35,angle=0]{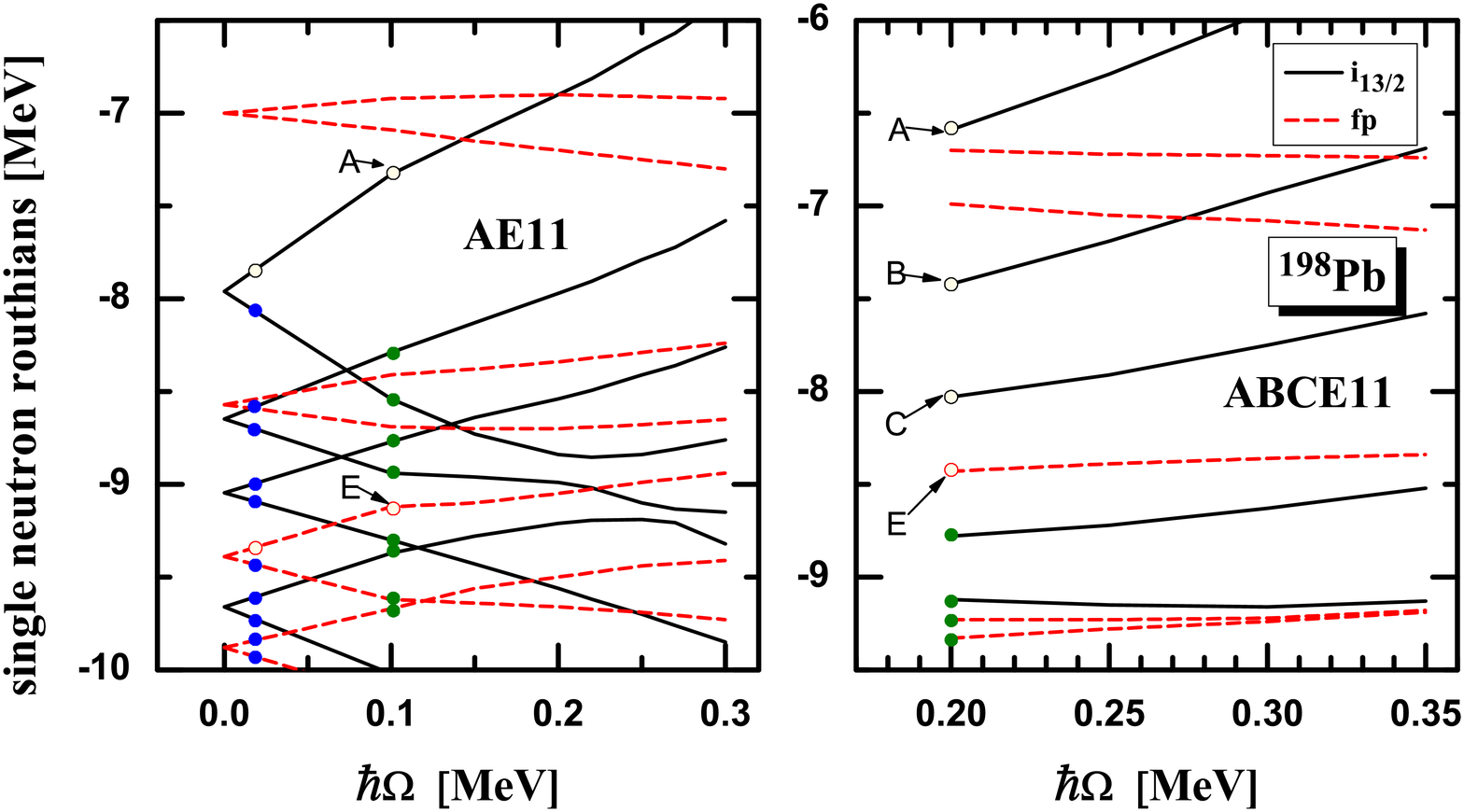}
}\caption{(Color online) Single-particle Routhians for the neutrons in $^{198}$Pb as a function of the rotational frequency based on the configurations AE11 and ABCE11. The blue dots indicate the occupied levels at $\Omega=0$ and the green dots indicate the occupied levels at the band heads with the configuration AE11 (left panel) and ABCE11 (right panel). Further details are given in the text.}
\label{fig1}
\end{figure*}

The magnetic rotational bands investigated in the present work include bands 1 and 3 in the nucleus $^{198}$Pb~\cite{Clark1993121} and bands 1 and 2 in $^{199}$Pb~\cite{Baldsiefen1994521}. In both nuclei we have the same proton configurations for these bands, i.e., $\pi[s_{1/2}^{-2}h_{9/2}i_{13/2}]11^-$. In Fig.~\ref{fig1} we show the single-particle Routhians for the neutrons in the nucleus $^{198}$Pb as a function of the rotational frequency $\Omega$ for the two configurations AE11 (left panel) and ABCE11 (right panel). The blue dots indicate the occupied levels at $\Omega=0$ and the green dots indicate the occupied levels at the band head of the bands with the configurations AE11 (left panel) and ABCE11 (right panel). The positive parity levels belonging to the $\nu i_{13/2}$ orbit are given by full black curves and the neutron levels with negative parity $(pf)$ are indicated by dashed red curves. In these bands, a backbending phenomenon has been observed caused by the alignment of a pair of $i_{13/2}$ neutrons.  Before the backbending, the neutron configurations  $\nu [i_{13/2}^{-1}(fp)^{-1}]$, $\nu i_{13/2}^{-2}$, $\nu i_{13/2}^{-1}$, and $\nu [i_{13/2}^{-2}(fp)^{-1}]$ have been assigned to bands 1 and 3 in $^{198}\rm Pb$ and to bands 1 and 2 in $^{199}\rm Pb$, respectively. After the backbending, they  become $\nu[ i_{13/2}^{-3}(fp)^{-1}]$, $\nu i_{13/2}^{-4}$, $\nu i_{13/2}^{-3}$, and $\nu[ i_{13/2}^{-4}(fp)^{-1}]$~\cite{Amita2000AtomicDataandNuclearDataTables, NPA683(2001)108}. As in Ref.~\cite{Baldsiefen1994521} a short-hand notation is used for these configurations: A, B, C, and D denote $\nu i_{13/2}$ holes with positive parity and  E denotes a neutron hole with negative parity (mainly of $f_{5/2}$ and $p_{3/2}$ origin). The proton configuration $\pi[s_{1/2}^{-2}h_{9/2}i_{13/2}]11^-$ is abbreviated by its spin number 11. Therefore, the configurations presented above are referred as AE11, AB11, A11, and ABE11 for the bands before the backbending, and as ABCE11, ABCD11, ABC11, and ABCDE11 after the backbending.

The calculations of this work have been carried out with the covariant point-coupling density functional PC-PK1~\cite{Zhao2010Phys.Rev.C}, while pairing correlations are neglected. The Dirac equation~[Eq. (\ref{dirac})] of the nucleons is solved in a three-dimensional harmonic oscillator basis in Cartesian coordinates, introduced in Ref.~\cite{KR.88} and discussed in detail in Ref.~\cite{Peng2008Phys.Rev.C}. In the present calculations, $N_f=12$ major oscillator shells are used. This provides a satisfactory accuracy of the results. By increasing $N_f$ from $12$ to $14$, we find for the ground state of the nucleus $^{198}\rm Pb$ changes of less than $0.04\%$ for the total energies and less than $5\%$ for the mass quadrupole moments. It has also been checked that in the solutions of the present cranking calculations the direction of the cranking axis $\bm\Omega$ and the direction of angular-momentum axis  $\bm J$ are identical, which means that self-consistency has been achieved.

\section{RESULTS AND DISCUSSION}
\label{section3}
\begin{figure*}
\centerline{
\includegraphics[scale=0.35,angle=0]{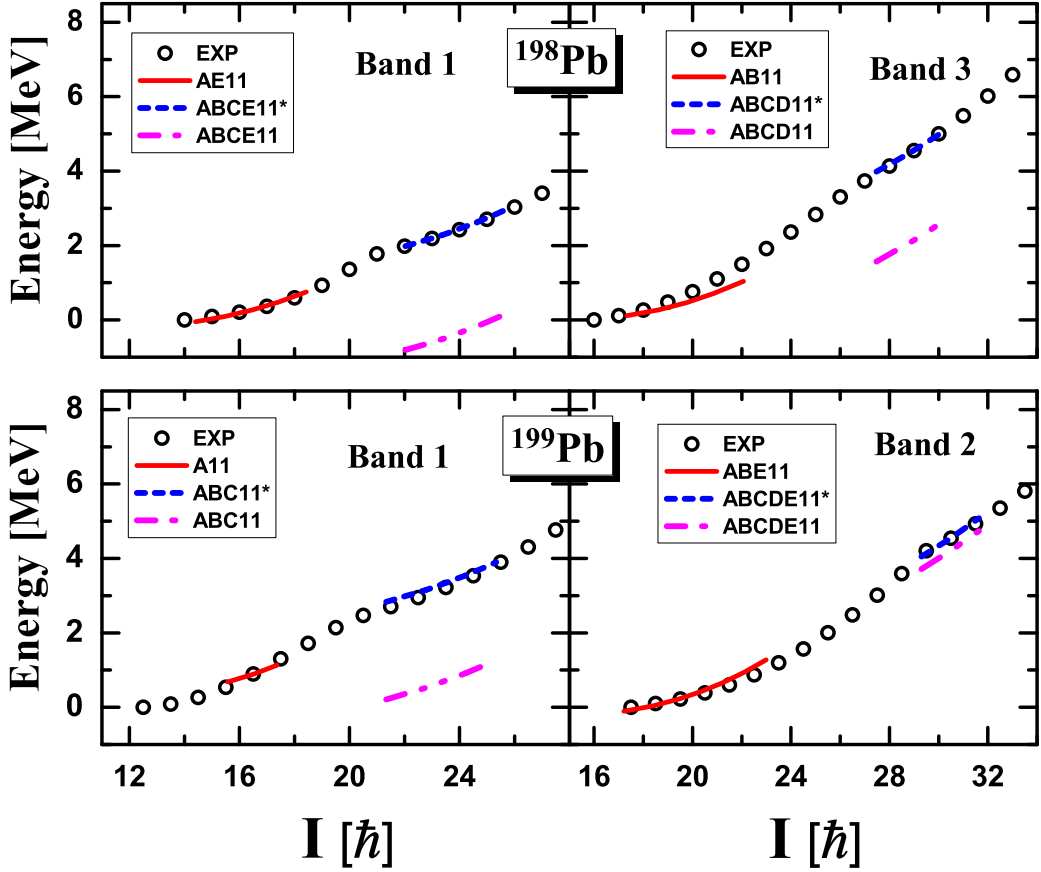}
} \caption{(Color online)
Energy spectra in the TAC-RMF calculations compared with the data~\cite{NPA683(2001)108,EPJA5(1999)257} for bands 1 and 3 in $^{198}\rm Pb$ (upper panels) and bands 1 and 2 in $^{199}\rm Pb$ (lower panels). The energies at $I= 15\hbar$, $17\hbar$, $33/2\hbar$, and $39/2\hbar$ are taken as references for the bands 1 and 3 in $^{198}\rm Pb$ and bands 1 and 2 in $^{199}\rm Pb$, respectively. Energies for the configurations ABCE11* and ABCD11* in $^{198}\rm Pb$ as well as ABC11* and ABCDE11* in $^{199}\rm Pb$ are renormalized to the energies at $I=22\hbar$, $30\hbar$, $51/2\hbar$, and $61/2\hbar$, respectively.}
\label{fig2}
\end{figure*}

In Fig.~\ref{fig2} we show the calculated energy spectra for the bands 1 and 3 in $^{198}\rm Pb$ and for the bands 1 and 2 in $^{199}\rm Pb$ in comparison with the experimental data of Refs.~\cite{NPA683(2001)108,EPJA5(1999)257}. For certain regions of angular momenta, the calculated values are missing, as, for instance, $I=19-21\hbar$ in band 1 of $^{198}\rm Pb$.
This is because, as discussed in Ref.~\cite{Peng2008Phys.Rev.C}, due to the level crossing connected with the backbending phenomenon, we could not find converged solutions for these angular-momentum values. It can be seen that the present TAC-RMF calculations reproduce well the experimental energies for all the four bands but underestimate the particle-hole excitation energies at the band head of the configurations ABCE11 and ABCD11 in $^{198}\rm Pb$ as well as ABC11 and ABCDE11 in $^{199}\rm Pb$. In comparison with the PQTAC calculations~\cite{Frauendorf1993Nucl.Phys.,Chmel2007Phys.Rev.C}, these underestimations can be explained by the pairing correlations and will be further investigated in the future. At the moment, we compensate for these underestimations by choosing different references for the configurations involved.

\begin{figure*}
\centerline{
\includegraphics[scale=0.35,angle=0]{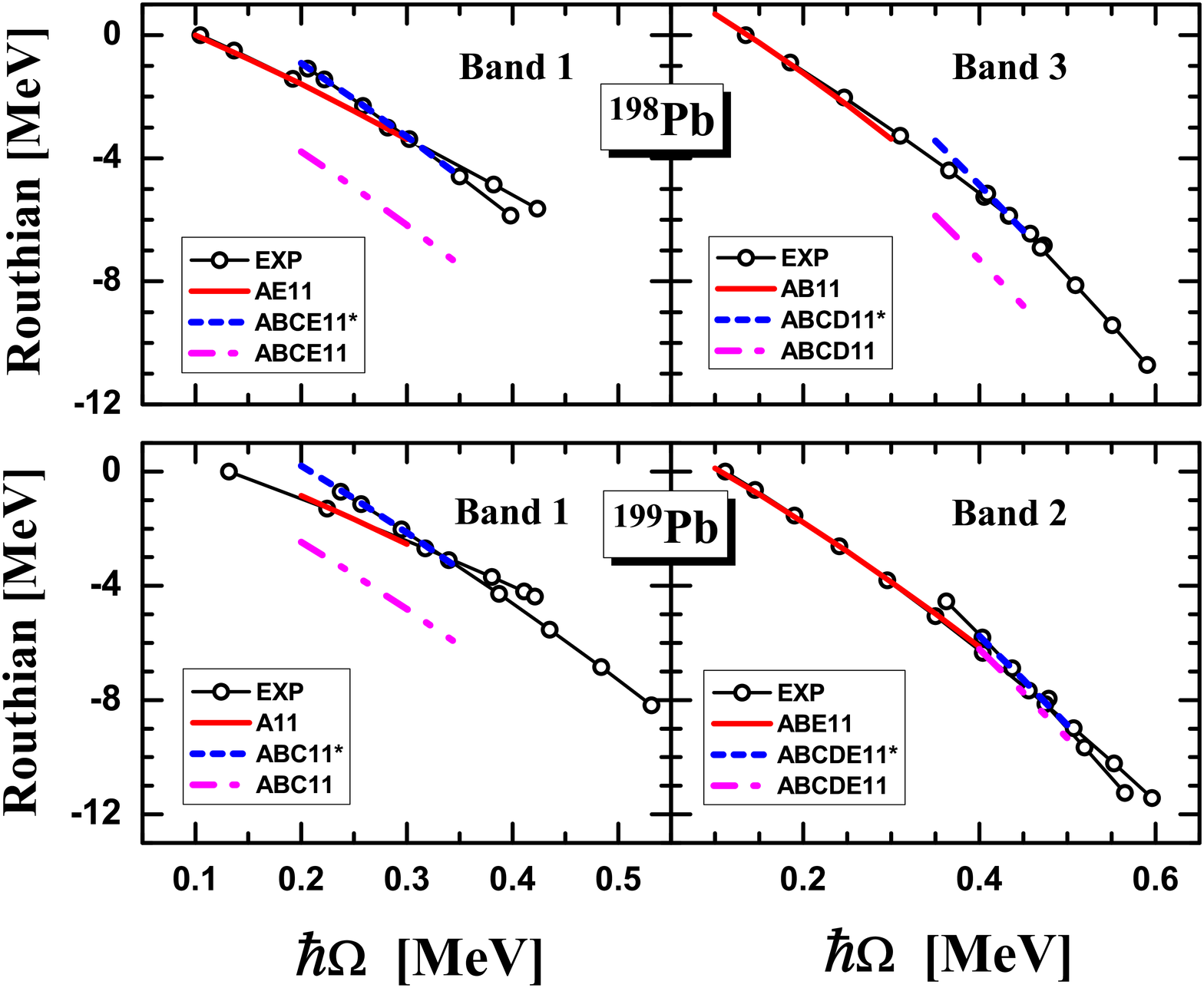}
} \caption{(Color online) Calculated total Routhians as functions of the rotational frequency in compared with the data~\cite{NPA683(2001)108,EPJA5(1999)257} for bands 1 and 3 in $^{198}\rm Pb$ (upper panels) and bands 1 and 2 in $^{199}\rm Pb$ (lower panels). The values at $\hbar\Omega= 0.1$, 0.15, 0.25, and 0.1 MeV are taken as references for the bands 1 and 3 in $^{198}\rm Pb$ and bands 1 and 2 in $^{199}\rm Pb$, respectively. Routhians for the configurations ABCE11*, ABCD11*, ABC11*, and ABCDE11* are renormalized to the values at $\hbar\Omega= 0.2$, 0.4, 0.25, and 0.4 MeV, respectively.}
\label{fig3}
\end{figure*}

In Fig.~\ref{fig3} we show the calculated total Routhians for the bands 1 and 3 in $^{198}\rm Pb$ and for the bands 1 and 2 in $^{199}\rm Pb$ as functions of the rotational frequency in comparison with the data of Refs.~\cite{NPA683(2001)108,EPJA5(1999)257}. It can be seen that the present TAC-RMF calculations reproduce well the experimental total Routhians for all the four bands before band crossing. However, one could see that the band crossings appear at too low frequency due to the neglect of pairing. In the PQTAC results, which include pairing, it could get proper band crossing frequency~\cite{Frauendorf1993Nucl.Phys.}. This is achieved in the present work by renormalizing the total Routhians after band crossing.

\begin{figure*}
\centerline{
\includegraphics[scale=0.35,angle=0]{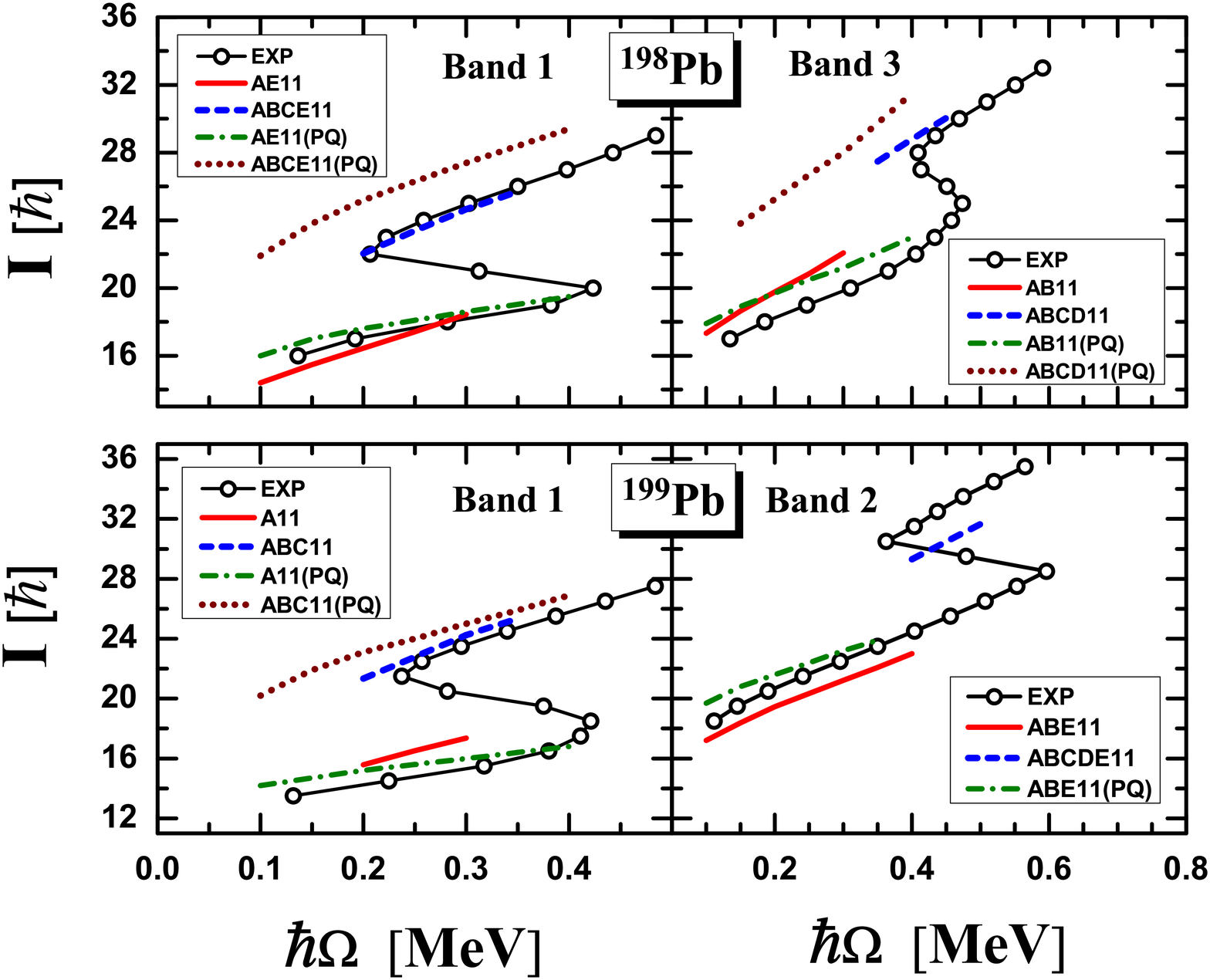}
} \caption{(Color online) Angular momenta as  functions of the
rotational frequency in the TAC-RMF calculations
compared with the data~\cite{NPA683(2001)108,EPJA5(1999)257} and the PQTAC results~\cite{Chmel2007Phys.Rev.C} for bands 1 and 3  in $^{198}\rm Pb$ (upper panels) and
bands 1 and 2  in $^{199}\rm Pb$ (lower panels). The configurations with ``(PQ)'' denote the corresponding
results of PQTAC calculations.}
\label{fig4}
\end{figure*}

The experimental rotational frequency $\Omega_{\rm exp}$ is extracted from the energy spectra by the relation
 \begin{equation}
    \hbar\mathit\Omega_{\rm exp}=\frac{1}{2}[E_\gamma(I+1\rightarrow I)+E_\gamma(I\rightarrow I-1)]\approx\frac{dE}{dI}.
 \end{equation}
 In Fig.~\ref{fig4}, the calculated total angular momenta of the bands 1 and 3 in $^{198}\rm Pb$ and the bands 1 and 2 in $^{199}\rm Pb$ as functions of the rotational frequency are shown in comparison with the experimental data~\cite{NPA683(2001)108,EPJA5(1999)257} and the PQTAC results~\cite{Chmel2007Phys.Rev.C}.
 It is found that both the TAC-RMF and the PQTAC results agree well with the experimental data. This shows that the TAC calculations can reproduce the relative changes of moment of inertia within the different bands rather well. In both nuclei the TAC calculations support that the backbendings arise through an excitation of a neutron-hole pair in the $i_{13/2}$ shell, i.e., by the transitions in the configurations AE11$\rightarrow$ABCE11 in band 1 of $^{198}\rm Pb$,  AB11$\rightarrow$ABCD11 in band 3 of $^{198}\rm Pb$,  A11$\rightarrow$ABC11 in band 1 of $^{199}\rm Pb$, and  ABE11$\rightarrow$ABCDE11 in band 2 of $^{199}\rm Pb$. In detail, before the backbending the spins values found in the TAC-RMF and PQTAC models differ from experimental values up to $2\hbar$. After the backbending, the phenomenological PQTAC results for the bands 1 and 3 in $^{198}\rm Pb$ are nearly $3\hbar$ larger than the experimental values and the TAC-RMF results. Comparing with the experimental values in Fig.~\ref{fig4}, the appearance of backbending is clearly seen for each band.

\begin{figure*}
\centerline{
\includegraphics[scale=0.35,angle=0]{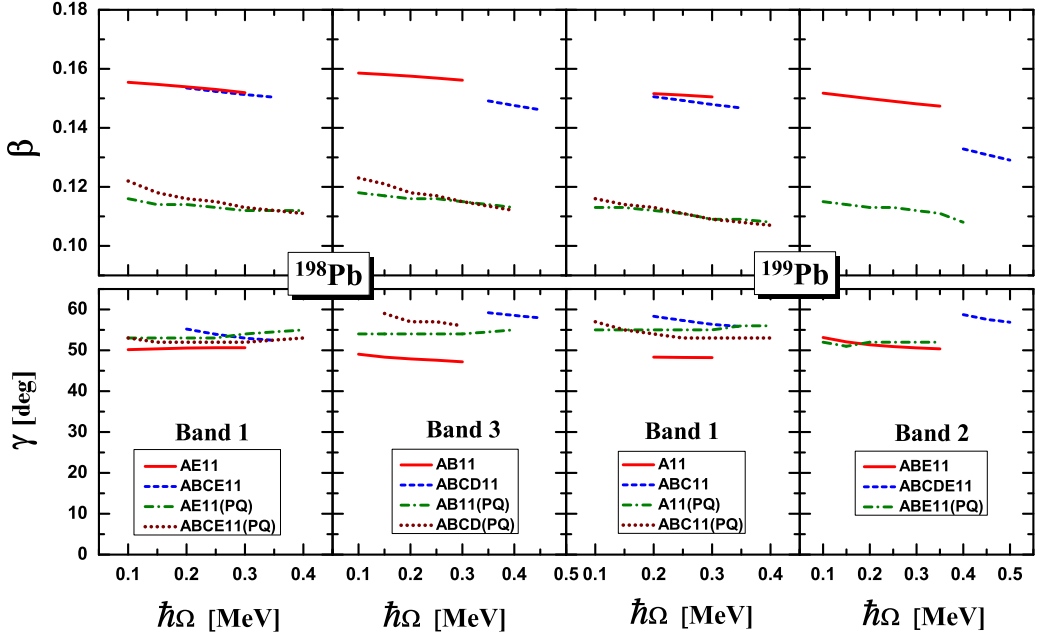}
} \caption{(Color online) Deformation parameters $\beta$ (upper panels) and $\gamma$ (lower panels) as  functions of the rotational frequency in the TAC-RMF calculations
compared with  the PQTAC results~\cite{Chmel2007Phys.Rev.C} for bands 1 and 3  in $^{198}\rm Pb$ and
bands 1 and 2  in $^{199}\rm Pb$.}
\label{fig5}
\end{figure*}

In Fig.~\ref{fig5}, the deformation parameters $\beta$ and $\gamma$ obtained in the self-consistent TAC-RMF calculations are compared with the phenomenological PQTAC results~\cite{Chmel2007Phys.Rev.C}. In the TAC-RMF calculations, the quadrupole deformations are small and remain almost constant, mainly lying around $\beta = 0.15$ for each configuration.  The PQTAC calculations produce the same tendency but smaller deformations, typically around $\beta =0.11$. Meanwhile, the $\gamma$ values vary between $47^\circ$ and $59^\circ$ which means small triaxiality  close to oblate axial symmetry in the TAC-RMF calculations. This is consistent with the PQTAC results of Ref.~\cite{Chmel2007Phys.Rev.C}.

\begin{figure*}
\centerline{
\includegraphics[scale=0.35,angle=0]{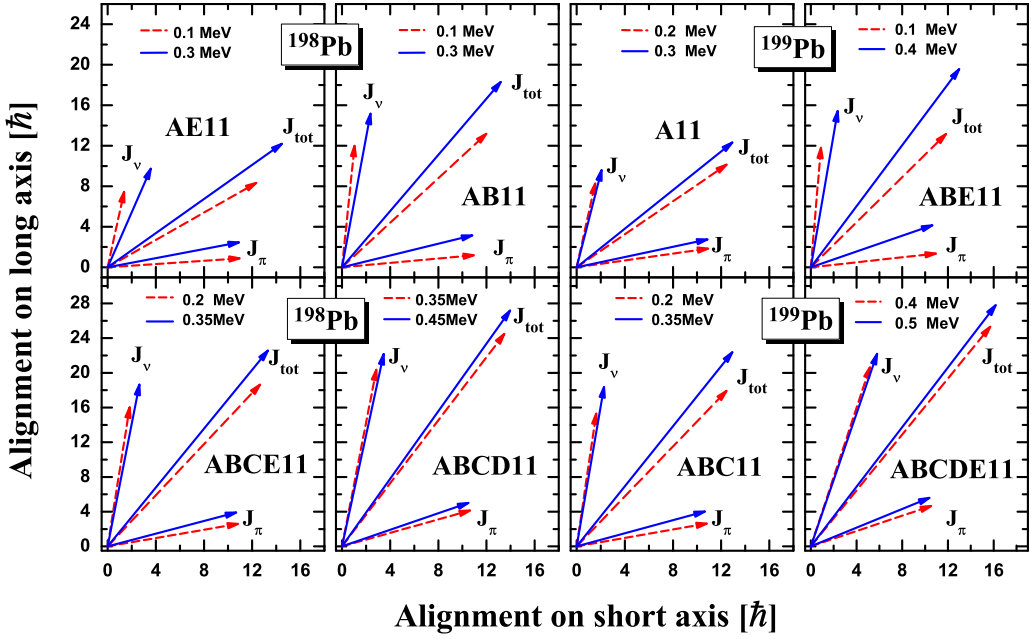}
} \caption{(Color online) Composition of the total angular momentum at both the minimum and the maximum
     rotational frequencies in TAC-RMF calculations for bands 1 and 3  in $^{198}\rm Pb$ and
bands 1 and 2  in $^{199}\rm Pb$. Upper (lower) panels are results for the rotation before (after) backbending.}
\label{fig6}
\end{figure*}

In order to examine the shears mechanism for the magnetic rotational bands in the nuclei $^{198}\rm Pb$ and $^{199}\rm Pb$, we show in Fig.~\ref{fig6} the proton and the neutron angular-momentum vectors $\bm J_\pi$ and $\bm J_\nu$ as well as the total angular-momentum vectors $\bm J_{\rm tot}=\bm J_\pi+\bm J_\nu$ at both the minimum and the maximum rotational frequencies in TAC-RMF calculations for the bands 1 and 3  in $^{198}\rm Pb$ and for the bands 1 and 2  in $^{199}\rm Pb$. The proton and neutron angular momenta $\bm J_\pi$ and $\bm J_\nu$ are defined as
\begin{equation}
    \bm J_\pi=\langle\bm{\hat J}_\pi\rangle=\sum_{p=1}^{Z}\langle p|\hat J|p\rangle, \quad\quad \bm J_\nu=\langle \bm{\hat J}_\nu\rangle=\sum_{n=1}^{N}\langle n|\hat J|n\rangle,
\end{equation}
where the sum runs over all the proton (or neutron) levels occupied in the cranking wave function in the intrinsic system.

For the magnetic dipole bands in $^{198}\rm Pb$ and $^{199}\rm Pb$, the contributions to the angular momenta come mainly from the high $j$ orbitals, i.e., the $i_{13/2}$ neutron (s) as well as $h_{9/2}$ and $i_{13/2}$ protons. At the band head, the proton particles excited across the closed $Z = 82$ shell gap into the $h_{9/2}$ and $i_{13/2}$ orbitals contribute to the proton angular momentum along the short axis, and the neutron hole(s) at the upper end of the $i_{13/2}$ shell contribute to the neutron angular momentum along the long axis. By comparing the upper panels (before backbending) with the lower ones (after backbending) in Fig.~\ref{fig6}, one finds that after the backbending the neutron angular-momentum vectors are considerably larger, because they contain the contributions of an aligned pair of $i_{13/2}$ neutron holes. For all the cases, the proton and neutron angular-momentum vectors form the two blades of the shears. As the frequency increases, the two blades move toward each other increase the larger angular momentum, while the direction of the total angular momentum stays nearly unchanged. In this way, the shears mechanism is clearly presented.

\begin{figure*}
\centerline{
\includegraphics[scale=0.35,angle=0]{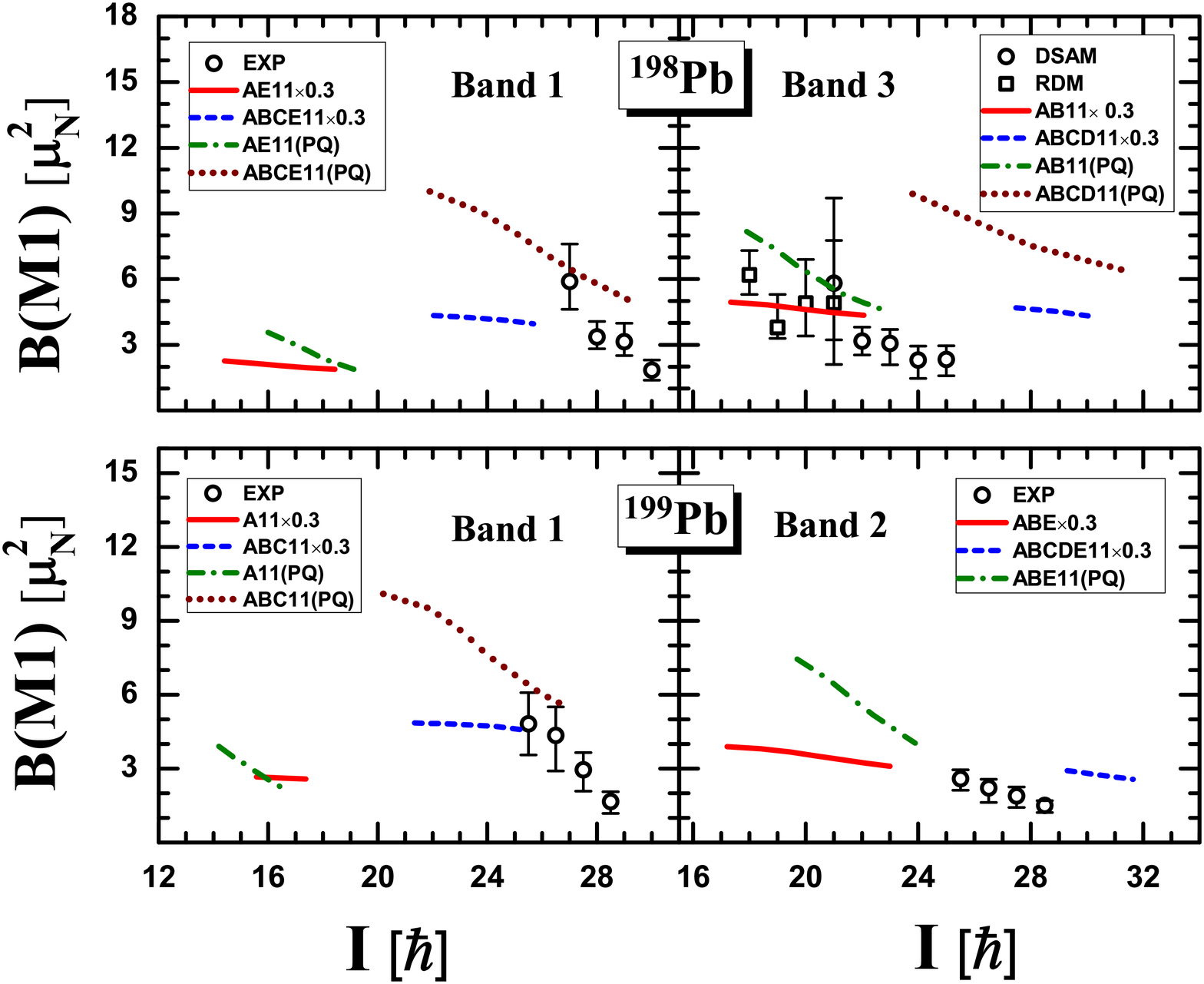}
}
\caption{(Color online) $B(M1)$ values as  functions of the total
angular momentum in the TAC-RMF calculations compared
with the data and the PQTAC results~\cite{Chmel2007Phys.Rev.C} for bands 1 and 3  in $^{198}\rm Pb$ (upper panels) and
bands 1 and 2  in $^{199}\rm Pb$ (lower panels). Circles and squares denote experimental data from DSAM~\cite{Clark1997Phys.Rev.Lett.} and RDM~\cite{PRC58(1998)1876}, respectively. 
}
\label{fig7}
\end{figure*}

A typical characteristic of magnetic rotation are strongly enhanced $M1$ transitions at low spins as well their decrease with increasing spin.
In Fig.~\ref{fig7} we show the calculated $B(M1)$ values as functions of the total angular momentum for the bands 1 and 3  in $^{198}\rm Pb$ and for
the bands 1 and 2  in $^{199}\rm Pb$ in comparison with the data~\cite{Clark1997Phys.Rev.Lett., PRC58(1998)1876} and the PQTAC results~\cite{Chmel2007Phys.Rev.C}.
The TAC-RMF calculations reproduce the decrease of the observed $B(M1)$ values with increasing spin. However, as it has been observed already in earlier calculations~\cite{Madokoro2000Phys.Rev.C, Zhao2011181} the absolute values show discrepancies. As shown in Fig.~\ref{fig7}, one has to attenuate the results by a factor 0.3 in order to reproduce the absolute $B(M1)$ values. The same factor has been used in Refs.~\cite{Madokoro2000Phys.Rev.C, Zhao2011181}. So far the origin of this attenuation factor is not understood in detail.
As discussed in Ref.~\cite{Zhao2011181}, there are, however, several reasons: (a) Pairing correlations strongly affect the levels in the neighborhood of the Fermi surface. This causes a strong reduction for the $B(M1)$ values with major contributions from the valence particles or holes. (b) The coupling to complex configurations such as particle vibrational coupling (Arima Horie effect~\cite{Arima1954PTP,Arima2011SCSG}) leads in all cases to a quenching of the $B(M1)$ values for neutron configurations~\cite{Bauer1973NPA,Matsuzaki1988PTP}. (c) Meson exchange currents and higher corrections also cause a reduction of the effective $g$ factors for the neutrons~\cite{Towner1987PR,Li2011PTP,Li2011SCSG}. However, it is not the absolute $B(M1)$ values, which characterize the shear bands, but rather the behavior of these values with increasing angular momentum.
On the other side, the absolute values of phenomenological PQTAC results agree with the observed $B(M1)$ data and the attenuated TAC-RMF results. However, they show a sharper decreasing trend as compared with the TAC-RMF calculations. The agreement between the calculated and experimental $B(M1)$ values and their trend shows a convincing confirmation of the shears mechanism.

\begin{figure*}
\centerline{
\includegraphics[scale=0.35,angle=0]{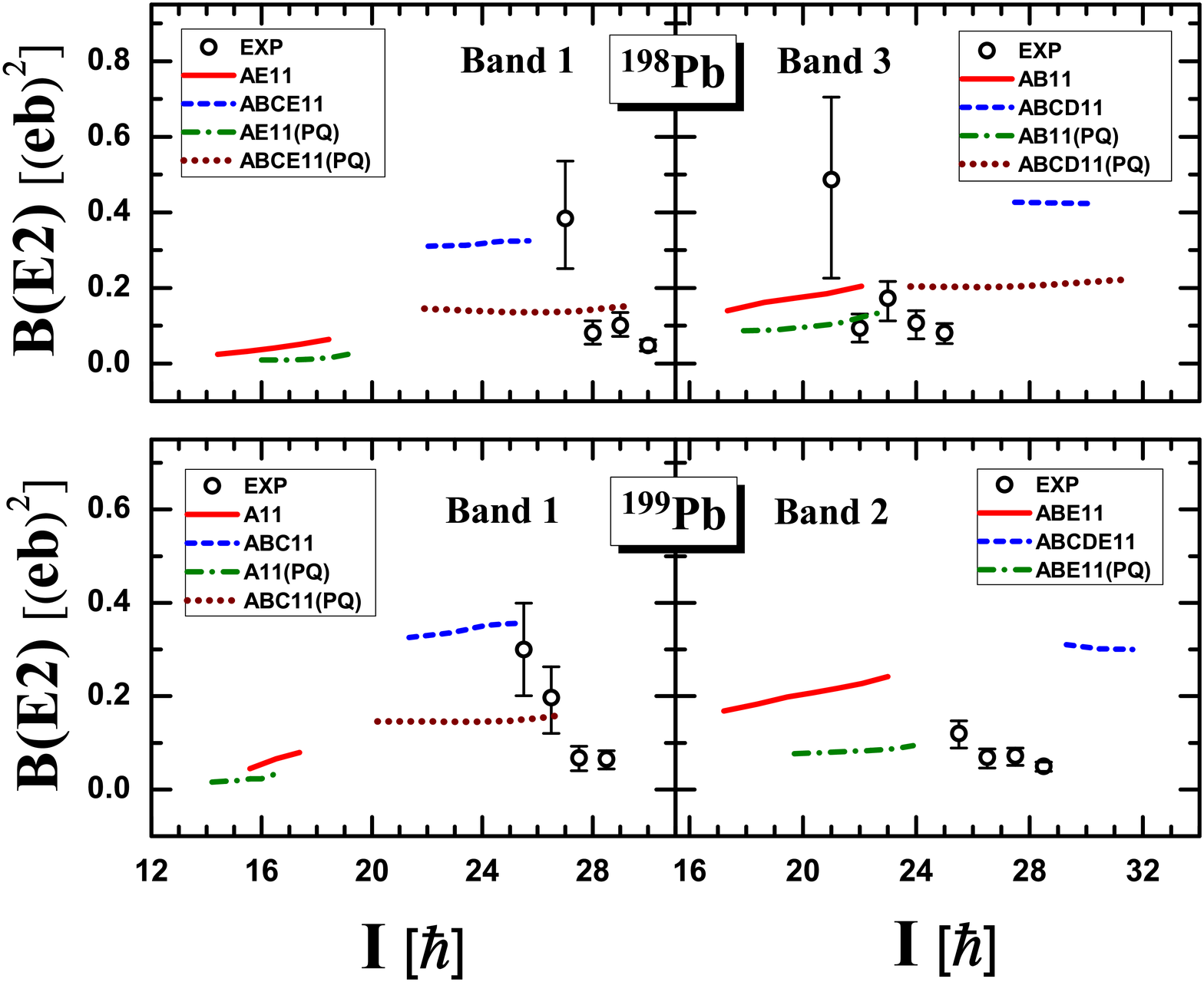}
}
\caption{(Color online) $B(E2)$ values as  functions of the total
angular momentum in the TAC-RMF calculations compared
with the data from DSAM~\cite{Clark1997Phys.Rev.Lett.} and the PQTAC results~\cite{Chmel2007Phys.Rev.C} for bands 1 and 3  in $^{198}\rm Pb$ (upper panels) and
bands 1 and 2  in $^{199}\rm Pb$ (lower panels). }
\label{fig8}
\end{figure*}

In contrast to the enhanced $M1$ transitions, the $E2$ transitions are very weak for magnetic rotational bands.
In Fig.~\ref{fig8} we show the calculated $B(E2)$ values as functions of the total angular momentum and compare them
with the DSAM-data of Ref.~\cite{Clark1997Phys.Rev.Lett.} and the theoretical PQTAC results of Ref.~\cite{Chmel2007Phys.Rev.C} for bands 1 and 3  in $^{198}\rm Pb$ and
bands 1 and 2  in $^{199}\rm Pb$. The $B(E2)$ values in the TAC-RMF calculations are in reasonable agreement
with the data and show a roughly constant trend. This is consistent with the calculated nearly constant quadrupole deformation in each configuration. Compared to the PQTAC results, the TAC-RMF calculations predict larger $B(E2)$ values, in accordance with the larger deformations shown in Fig.~\ref{fig4}.
For band 3 in $^{198}\rm Pb$, both the TAC-RMF and the PQTAC calculations give nearly constant or even slightly increasing $B(E2)$ values, which  differs from the results obtained within the framework of a geometrical approach~\cite{Clark2000Annu.Rev.Nucl.Part.Sci.}. 
More accurate experimental values are expected to clarify the spin dependence of the $B(E2)$ values.

\section{Conclusion}
\label{conclusion}

The fully microscopic and self-consistent TAC-RMF model based on the point-coupling density functional PC-PK1 has been applied to investigate the shears bands in the nuclei $^{198}\rm Pb$ and $^{199}\rm Pb$. This is the first fully microscopic investigation of magnetic dipole bands in this region. The energy spectra, the relation between spin and rotational frequency, deformation parameters, and reduced $M1$ and $E2$ transition probabilities have been calculated and compared with data and with the results obtained from phenomenological PQTAC models. So far, pairing correlations are neglected in the TAC-RMF model and, therefore, the band head had to be renormalized. The remaining energy spectra are well reproduced in the TAC-RMF calculations. Both the TAC-RMF and the PQTAC calculated angular momenta agree well with the experimental data for various configurations, showing that the TAC calculations can reproduce the moments of inertia rather well. Considering the transitions in configurations it is shown that the occurrence of the backbending is accompanied with the alignment of an $i_{13/2}$ neutron-hole pair. The decrease in the $B(M1)$ values calculated by the two TAC models is in good agreement with the experimental values. The fact that the $B(M1)$ values decrease with the increasing frequency indicates the appearance of the shears mechanism in $^{198, 199}$Pb. This mechanism can be also clearly seen in our calculations by decomposing the total angular momenta at different frequencies into contributions of protons and neutrons. The roughly constant trends of the calculated $B(E2)$ values and the quadrupole deformations for different rotational frequency are consistent with the $\rm PQTAC$ calculations but the self-consistent values are larger than the those deduced from the PQTAC calculations.

It should be mentioned that the time-odd terms play important roles in the description of magnetic rotation which has been discussed in details for the band based on the configuration $\pi h^2_{11/2} \otimes \nu h^{-2}_{11/2}$ in $^{142}$Gd~\cite{Peng2009CPL}. The present calculations give similar conclusions that the time-odd terms will lower the total Routhians and their influence is negligible for the $B(E2)$ values but appreciable for the $B(M1)$ values.

Following these first self-consistent calculations of magnetic dipole bands in the Pb region, further works are planed or in progress, e.g., the pairing correlations and their contribution to the excitation energies at the various band heads, systematic calculations with other covariant density functionals to clarify details of the deformation parameters and the transition probabilities, and more investigations of other nuclei in this area to understand the large variety of experimental data in this area.

{\center{\bf ACKNOWLEDGMENTS}}

This work was partly supported by the Major State 973 Program
2007CB815000, the National Natural Science Foundation of China
under Grants No. 10975007, No. 10975008, No. 11005069, and No. 11175002; the Research Fund for the Doctoral Program of Higher Education under Grant No. 20110001110087; and the Oversea Distinguished Professor Project from Ministry of Education No. MS2010BJDX001. We also acknowledge partial support from the DFG Cluster of Excellence \textquotedblleft Origin and Structure of the
Universe\textquotedblright\ (www.universe-cluster.de).



\newpage

\begin{thebibliography}{64}
\bibitem{Frauendorf1993Nucl.Phys.}
S. Frauendorf, Nucl. Phys. {\bf A557},  259c  (1993).

\bibitem{Frauendorf1994}
S. Frauendorf, J. Meng, and J. Reif,  in {\em Report
  LBL35687}, Vol.~II, edited by M.~A. Deleplanque (University of California Press, Berkeley, 1994),
  p.\ 52.

\bibitem{Frauendorf1996}
S. Frauendorf, in {\em Proceedings of Workshop on Gammasphere Physics, Berkeley, 1995}, edited by M.~A. Deleplanque (World Scientific, Singapore, 1996),
  p.\ 272.

\bibitem{Frauendorf1997Z.Phys.A}
S. Frauendorf, Z. Phys. {\bf A358},  163  (1997).

\bibitem{RMP73(2001)463}
S. Frauendorf, Rev. Mod. Phys. {\bf 73},  463  (2001).


\bibitem{Amita2000AtomicDataandNuclearDataTables}
Amita, A.~K. Jain, and B. Singh, At. Data Nucl. Data Tables {\bf 74},  283
  (2000).

\bibitem{Torres2008Phys.Rev.C}
D.~A. Torres~{\it et al}, Phys. Rev. {\bf C78},  054318
   (2008).

\bibitem{Clark1992Phys.Lett.}
R.~M. Clark~{\it et al}, Phys. Lett. {\bf B275},  247  (1992).

\bibitem{Baldsiefen1992Phys.Lett.}
G. Baldsiefen~{\it et al}, Phys. Lett. {\bf B275},  252  (1992).


\bibitem{Kuhnert1992Phys.Rev.C}
A. Kuhnert~{\it et al}, Phys. Rev. {\bf C46},  133  (1992).

\bibitem{Clark1997Phys.Rev.Lett.}
R.~M. Clark~{\it et al}, Phys. Rev. Lett. {\bf 78},  1868  (1997).

\bibitem{PRC58(1998)1876}
R. Kr{\"u}cken~{\it et al}, Phys. Rev. {\bf C58},  R1876  (1998).

\bibitem{Frauendorf2000Nucl.Phys.}
S. Frauendorf, Nucl. Phys. {\bf A677},  115  (2000).


\bibitem{Frauendorf1996Nucl.Phys.}
S. Frauendorf, J. Reif, and G. Winter, Nucl. Phys. {\bf A601},  41  (1996).

\bibitem{PhysRevC.74.044310}
B.~G. Carlsson and I. Ragnarsson, Phys. Rev. {\bf C74},  044310  (2006).

\bibitem{Frauendorf1996ZP}
S. Frauendorf and J. Meng, Z. Phys. {\bf A356},  263  (1996).



\bibitem{Frauendorf1997NP}
S. Frauendorf and J. Meng, Nucl. Phys. {\bf A617},  131  (1997).

\bibitem{Chmel2007Phys.Rev.C}
S. Chmel, S. Frauendorf, and H. H{\"u}bel, Phys. Rev. {\bf C75},  044309
  (2007).

\bibitem{Olbratowski2002Acta.Phys.Pol.B}
P. Olbratowski, J. Dobaczewski, J. Dudek, T. Rzaca-Urban, Z. Marcinkowska, and
  R.~M. Lieder, Acta Phys. Pol. {\bf B33},  389  (2002).

\bibitem{Madokoro2000Phys.Rev.C}
H. Madokoro, J. Meng, M. Matsuzaki, and S. Yamaji, Phys. Rev. {\bf C62},
  061301(R)  (2000).

\bibitem{Peng2008Phys.Rev.C}
J. Peng, J. Meng, P. Ring, and S.~Q. Zhang, Phys. Rev. {\bf C78},  024313
  (2008).




\bibitem{Zhao2011181}
P.~W. Zhao, S.~Q. Zhang, J. Peng, H.~Z. Liang, P. Ring, and J. Meng, Phys.
  Lett. {\bf B699},  181  (2011).

\bibitem{Ring1996Prog.Part.Nucl.Phys.}
P. Ring, Prog. Part. Nucl. Phys. {\bf 37},  193  (1996).

\bibitem{Vretenar2005101}
D. Vretenar, A.~V. Afanasjev, G.~A. Lalazissis, and P. Ring, Phys. Rep. {\bf
  409},  101  (2005).

\bibitem{Meng2006Prog.Part.Nucl.Phys.}
J. Meng, H. Toki, S. G. Zhou, S. Q. Zhang, W. H. Long, and L. S. Geng, Prog.
  Part. Nucl. Phys. {\bf 57},  470  (2006).

\bibitem{KR.89}
W. Koepf and P. Ring, Nucl. Phys. {\bf A493},  61  (1989).




\bibitem{AA.10}
A.~V. Afanasjev and H. Abusara, Phys. Rev. {\bf C81},  014309  (2010).

\bibitem{HR.88}
U. Hofmann and P. Ring, Phys. Lett. {\bf B214},  307  (1988).


\bibitem{PRC74(2006)024307}
J.~M. Yao, H. Chen, and J. Meng, Phys. Rev. {\bf C74},  024307  (2006).

\bibitem{ARK.00}
A.~V. Afanasjev, P. Ring, and J. K{\"o}nig, Nucl. Phys. {\bf A676},  196
  (2000).

\bibitem{DD.95}
J. Dobaczewski and J. Dudek, Phys. Rev. {\bf C52},  1827  (1995).

\bibitem{DT.56}
H.~P. D{\"u}rr and E. Teller, Phys. Rev. {\bf 101},  494  (1956).



\bibitem{Due.56}
H.~P. D{\"u}rr, Phys. Rev. {\bf 103},  469  (1956).

\bibitem{Wal.74}
J.~D. Walecka, Ann. Phys. (N.Y.) {\bf 83},  491  (1974).

\bibitem{SW.86}
B.~D. Serot and J.~D. Walecka, Adv. Nucl. Phys. {\bf 16},  1  (1986).

\bibitem{BB.77}
J. Boguta and A.~R. Bodmer, Nucl. Phys. {\bf A292},  413  (1977).

\bibitem{NL3}
G.~A. Lalazissis, J. K{\"o}nig, and P. Ring, Phys. Rev. {\bf C55},  540
  (1997).

\bibitem{TP.05}
B.~G. Todd-Rutel and J. Piekarewicz, Phys. Rev. Lett. {\bf 95},  122501
  (2005).

\bibitem{PK1}
W.~H. Long, J. Meng, N. Van~Giai, and S.-G. Zhou, Phys. Rev. {\bf C69},  034319
   (2004).

\bibitem{BT.92}
R. Brockmann and H. Toki, Phys. Rev. Lett. {\bf 68},  3408  (1992).

\bibitem{TW.99}
S. Typel and H.~H. Wolter, Nucl. Phys. {\bf A656},  331  (1999).

\bibitem{DD-ME1}
T. Nik{\v{s}}i{\'{c}}, D. Vretenar, P. Finelli, and P. Ring, Phys. Rev. {\bf
  C66},  024306  (2002).




\bibitem{DD-ME2}
G.~A. Lalazissis, T. Nik{\v{s}}i{\'{c}}, D. Vretenar, and P. Ring, Phys. Rev.
  {\bf C71},  024312  (2005).

\bibitem{HKL.01}
F. Hofmann, C.~M. Keil, and H. Lenske, Phys. Rev. {\bf C64},  034314  (2001).

\bibitem{SOA.05}
M. Serra, T. Otsuka, Y. Akaishi, P. Ring, and S. Hirose, Prog. Theor. Phys.
  {\bf 113},  1009  (2005).

\bibitem{HSR.07}
S. Hirose, M. Serra, P. Ring, T. Otsuka, and Y. Akaishi, Phys. Rev. {\bf C75},
  024301  (2007).

\bibitem{Ing.56}
D.~R. Inglis, Phys. Rev. {\bf 103},  1786  (1956).




\bibitem{RBM.70}
P. Ring, R. Beck, and H.~J. Mang, Z. Phys. {\bf 231},  10  (1970).

\bibitem{RM.74}
P. Ring and H.~J. Mang, Phys. Rev. Lett. {\bf 33},  1174  (1974).

\bibitem{FMR.79}
J. Fleckner, U. Mosel, P. Ring, and H.~J. Mang, Nucl. Phys. {\bf A331},  288
  (1979).

\bibitem{BFH.87}
P. Bonche, H. Flocard, and P.-H. Heenen, Nucl. Phys. {\bf A467},  115  (1987).

\bibitem{ERo.93}
J.~L. Egido and L.~M. Robledo, Phys. Rev. Lett. {\bf 70},  2876  (1993).




\bibitem{GDB.94}
A. Girod, J.~P. Delaroche, J.~P. Berger, and J. Libert, Phys. Lett. {\bf B325},
   1  (1994).

\bibitem{AKR.96}
A.~V. Afanasjev, J. K{\"o}nig, and P. Ring, Nucl. Phys. {\bf A608},  107
  (1996).

\bibitem{BMR.70}
R. Beck, H.~J. Mang, and P. Ring, Z. Phys. {\bf 231},  26  (1970).

\bibitem{Bengtsson1987}
H. Frisk and R. Bengtsson, Phys. Lett. {\bf B196}, 14 (1987)

\bibitem{Olbratowski2004Phys.Rev.Lett.}
P. Olbratowski, J. Dobaczewski, J. Dudek and W. P{\l}\'{o}ciennik, Phys. Rev. Lett. {\bf93}, 052501 (2004).


\bibitem{PRC46(1992)1757}
B.~A. Nikolaus, T. Hoch, and D.~G. Madland, Phys. Rev. {\bf C46},  1757  (1992).

\bibitem{Zhao2010Phys.Rev.C}
P.~W. Zhao, Z.~P. Li, J.~M. Yao, and J. Meng, Phys. Rev. {\bf C82},  054319
  (2010).

\bibitem{Zhao2011PRL}
P.~W. Zhao, J. Peng, and H.~Z. Liang, P. Ring, and J. Meng, Phys. Rev. Lett. {\bf107}, 122501 (2011)

\bibitem{NPA683(2001)108}
A. A.~G{\"o}rgen~{\it et al}, Nucl. Phys. {\bf A683},  108  (2001).

\bibitem{EPJA5(1999)257}
W. W.~Pohler~{\it et al}, Eur. Phys. J. {\bf A5},  257  (1999).

\bibitem{Clark1993121}
R.~M. Clark~{\it et al}, Nucl. Phys. {\bf A562},  121  (1993).

\bibitem{Baldsiefen1994521}
G. Baldsiefen~{\it et al}, Nucl. Phys. {\bf A574},  521  (1994).

\bibitem{KR.88}
W. Koepf and P. Ring, Phys. Lett. {\bf B212},  397  (1988).

\bibitem{Arima1954PTP}
A. Arima and H. Horie, Prog. Theor. Phys. {\bf11}, 509 (1954).

\bibitem{Arima2011SCSG}
A. Arima, Sci. China Ser. G - Phys. Mech. Astron. {\bf54}, 188 (2011).

\bibitem{Bauer1973NPA}
R. Bauer, J. Speth, V. Klemt, P. Ring, E. Werner, and T. Yamazaki, Nucl. Phys. {\bf A209}, 535 (1973).

\bibitem{Matsuzaki1988PTP}
M. Matsuzaki, Y. R. Shimizu, and K. Matsuyanagi, Prog. Theor. Phys. {\bf79}, 836 (1988).

\bibitem{Towner1987PR}
I.~S. Towner, Phys. Rep. {\bf155}, 263 (1987).

\bibitem{Li2011PTP}
J. Li, J.~M. Yao, J. Meng, and A. Arima, Prog. Theor. Phys. {\bf125}, 1185 (2011).

\bibitem{Li2011SCSG}
J. Li, J. Meng, P. Ring, J.~M. Yao, and A. Arima, Sci. China Ser. G - Phys. Mech. Astron. {\bf54}, 204 (2011).


\bibitem{Clark2000Annu.Rev.Nucl.Part.Sci.}
R.~M. Clark and A.~O. Macchiavelli, Annu. Rev. Nucl. Part. Sci. {\bf 50}, 1 (2000).

\bibitem{Peng2009CPL}
J. Peng and L. F. Xing, Chin. Phys. Lett. {\bf26}, 032101 (2009).

\end{thebibliography}
\end{document}